\documentclass{article}
\usepackage{graphicx}
\begin{document}


\centerline{\today \hfill{UCD-HEP-99-14}}
\centerline{\hfill { FERMILAB-Pub-99/245-T} }

\begin{center}
{\bf Drell-Yan and Diphoton Production at Hadron Colliders \\and
Low Scale Gravity Model} \\
Kingman Cheung \\
{\it Department of Physics, University of California, Davis, CA 95616 USA}\\

Greg Landsberg \\
{\it Department of Physics, Box 1843, Brown University, Providence, 
Rhode Island 02912-1843 USA}
\end{center}

\begin{abstract}
In the model of Arkani-Hamed, Dimopoulos, and Dvali where gravity is
allowed to propagate in the extra dimensions of very large size,
virtual graviton exchange between the standard model particles can
give rise to signatures that can be tested in collider experiments. We
study these effects in dilepton and diphoton production at hadron
colliders.  Specifically, we examine the double differential
cross-section in the invariant mass and scattering angle, which is
found to be useful in separating the gravity effects from the standard
model. In this work, sensitivity obtained using the double
differential cross-section is higher than that in previous
studies based on single differential distributions. Assuming no
excess of events over the standard model predictions, we obtain the
following 95\% confidence level lower limits on the effective Planck
scale: $0.9-1.5$ TeV in the Tevatron Run I, $1.3-2.5$ TeV in Run IIa,
$1.7-3.5$~TeV in Run IIb, and $6.5-12.8$ TeV at the LHC. The range of
numbers corresponds to the number of extra dimensions $n=7-2$.
\end{abstract}

\section{Introduction}

Recent advances in string theory have revolutionized particle phenomenology.
Namely, the previously unreachable Planck, string, and grand unification 
scales ($M_{\rm Pl}$, $M_{\rm st}$, and $M_{\rm GUT}$, respectively) can 
now be brought down to a TeV range \cite{tevs}.
If this is the case, one expects low energy phenomenology that can be
tested in current and future collider experiments.

An attractive realization of the above idea was recently proposed by
Arkani-Hamed, Dimopoulos, and Dvali \cite{ADD}.  In their model, the standard
model (SM) particles live on a D3-brane, predicted in the 
string theory, and the SM gauge interactions are confined to this brane. 
On the other hand, gravity is allowed to propagate in the extra
dimensions.  In order to bring the Planck scale ($10^{19}$ GeV) to the TeV 
range, 
the size of these compactified dimensions is made very large compared to 
$(M_{\rm weak})^{-1}$.  The relation among the Planck scale $M_{\rm Pl}$, 
size $R$ of
the extra dimensions, and the effective Planck scale $M_S$ is given by:
\begin{equation}
M^2_{\rm Pl} \sim M_S^{n+2} \; R^n \;,
\end{equation}
where $n$ is the number of extra (compactified) dimensions.   From 
this relation, the size $R$ of the compactified extra
dimensions can be estimated.  Assuming that the effective Planck scale
$M_S$ is in the TeV range,  it gives a very large $R$ of
the size of our solar system for $n=1$, which is obviously ruled out by the
experiment. However, for all $n\ge 2$ the expected $R$ is less than 
$1$~mm, and therefore does not contradict existing gravitational experiments.

With the SM particles residing on the brane and the graviton freely
propagating in the extra dimensions, the SM particles can couple to a
graviton with a strength comparable to that of the electroweak
interactions.  A graviton in the extra dimensions is equivalent, from
the 4D-point of view, to a tower of infinite number of Kaluza-Klein
(KK) states with masses $M_k = 2\pi k/R\;\; (k=0,1,2,...,\infty)$.
The coupling to each of these KK states is $\sim 1/M_{\rm Pl}$.
The overall coupling is, however, obtained by summing over all the KK
states, and thus is $\sim 1/M_S$. Since $M_S$ is in the TeV
range, the gravitational interaction is as strong as electroweak
interactions, and thus can give rise to many consequences that can be tested in
both the accelerator and non-accelerator experiments.

A large number of phenomenological studies in this area have recently
appeared.  Among these studies, the strongest lower bound on the
effective Planck scale (30--100 TeV for $n=2$) comes from 
astrophysical (SN1987A) and cosmological constraints~\cite{astro}.
Collider signals and constraints \cite{collider,me1,atwood,me2,eboli,mathews}
come from diboson, dilepton, dijet, top-pair production, and real 
graviton emissions.

In general, present collider 
experiments are sensitive to the effective Planck scale below $\sim 1$ TeV.
In Refs. \cite{me1,me2} the Drell-Yan and diphoton production at the 
Tevatron were used to
constrain the scale $M_S$. In these studies, however, only the invariant mass 
distribution of the lepton or photon pair is used.  We found that the
distribution in the central scattering angle, in addition to the 
invariant mass distribution, further helps to constrain the scale $M_S$.  

In this work, we use the double differential cross-section,
$d^2\sigma/d M d \cos\theta^*$, to probe the effective Planck
scale $M_S$ in Run I and Run II at the Tevatron and at the LHC.  
The advantage of using double differential distribution is that the
differences in the invariant mass and scattering angle between the SM 
and the gravity model can be contrasted simultaneously.
Furthermore, for a $2\to 2 $ process the invariant mass $M$ and the
central scattering angle $\cos\theta^*$ already span the entire phase
space.  We, therefore, do not need to optimize the kinematic cuts 
or choose optimal 
variables ({\it e.g.}, forward-backward asymmetry, charged forward-backward
asymmetry, etc.), because all the relevant information is already contained
in the $(M\times \cos\theta^*)$ distribution.  We will show that
sensitivity obtained in this study has improved substantially, 
compared to previous
studies, in which only single differential distributions were used. By
analyzing double differential distributions in dilepton and
diphoton production simultaneously, we are able to reach sensitivity 
on $M_S$ at the 95\% confidence level (C.L.)
as high as $0.9-1.5$ TeV in the Tevatron Run I, $1.3-2.5$ TeV in
Run IIa, $1.7-3.5$~TeV in Run IIb, and $6.5-12.8$ TeV at the LHC, for
$n=7-2$. This is our main result.

The organization of the paper is as follows. In the next section, we
give the cross section for dilepton and diphoton production in the
presence of strong TeV scale gravity. In Sec.~\ref{sec:method} 
we describe the procedures in estimating the
sensitivity limits. In Sec.~\ref{sec:results} we present our results
for the Tevatron and for the LHC, and we conclude in
Sec.~\ref{sec:conclusions}  

\section{Drell-Yan and Diphoton Production}
\label{sec:cs}

An effective Lagrangian
for the low scale gravity interactions between the SM particles and the
graviton was derived by Han {\it et al.} in Ref. \cite{collider}.  
This effective theory is valid up to a scale of about $M_S$. 
The Drell-Yan production, including the contributions from the SM, gravity,
and the interference terms, is given by \cite{me1}:
\begin{eqnarray}
\label{dilepton}
\frac{d^3\sigma}{dM_{\ell\ell} dy dz } &=& K \Biggr\{  \sum_q
\frac{M_{\ell\ell}^3}{192 \pi s} \;
f_q(x_1) f_{\bar q}(x_2)\; \Biggr [
(1+z)^2 \left( |M_{LL}|^2 + |M_{RR}|^2 \right )+ 
(1-z)^2 \left( |M_{LR}|^2 + |M_{RL}|^2  \right ) \nonumber \\
&&+
 \pi^2 \; \left( \frac{\cal F}{M_S^4} \right )^2 \; M_{\ell\ell}^4 \;
 (1-3z^2 + 4z^4 ) -
 8 \pi e^2 Q_e Q_q \left( \frac{\cal F}{M_S^4} \right ) z^3 \nonumber \\
&&+
\frac{8 \pi e^2 }{\sin^2 \theta_{\rm w} \cos^2\theta_{\rm w}} \; 
\left( \frac{\cal F}{M_S^4} \right ) \; 
\frac{M_{\ell\ell}^2}{ M_{\ell\ell}^2 - M_Z^2} \;
\left( g_a^e g_a^q \frac{1 - 3z^2}{2} - g_v^e g_v^q z^3 \right ) \;
\Biggr ]  \nonumber \\
&+& 
\frac{\pi M_{\ell\ell}^7}{128 s }  \; f_g(x_1) f_g(x_2) \;
\left( \frac{\cal F}{M_S^4} \right )^2 \; 
(1 - z^4 ) \; \Biggr\},
\end{eqnarray}
where
\begin{eqnarray}
M_{\alpha\beta} &=& \frac{e^2 Q_e Q_q}{\hat s} + 
\frac{e^2}{\sin^2\theta_{\rm w} \cos^2 \theta_{\rm w} } \;
\frac{g_\alpha^e g_\beta^q}{\hat s - M_Z^2} \;, \;\; \alpha,\beta =L,R 
\nonumber \\
\label{FF}
{\cal F} &=& \Biggr \{ 
\begin{array}{l}
\log \left( \frac{M_S^2}{\hat s} \right ) \;\; {\rm for}\;\; n=2 \;, \\
\frac{2}{n-2} \;\;\;\;\;\;\;\;\;\;\; {\rm for}\;\; n>2 \;.
\end{array}
\end{eqnarray}
Here $\sqrt{s}$ is the center-of-mass energy of the $p \bar p$ collision,
$z=\cos\theta^*$ is cosine of the scattering angle in the parton
center-of-mass frame, $y$ is the rapidity of the lepton pair, 
$f_{q/g}(x)$ is the parton distribution function,  and we have assumed that 
$M_S^2 \gg \hat s, |\hat t|, |\hat u|$.  
In the above equations, $\hat s = M_{\ell\ell}^2$, 
$x_{1,2} = \frac{M_{\ell\ell}}{\sqrt s} e^{\pm y}$,
$g_L^f = T_{3f} - Q_f \sin^2\theta_{\rm w}$,
$g_R^f = - Q_f \sin^2\theta_{\rm w}$,
$g_v^f = (g_L^f + g_R^f)/2$, and $g_a^f = (g_L^f - g_R^f)/2$.
It is implied that all possible $q\bar q$ initial states are summed over.
In what follows, we substitute  $\eta={\cal F}/M_S^4$  
for convenience and for use as a fit parameter.

Similarly, we calculate the diphoton production. The double differential
cross section is given by \cite{me2}:
\begin{eqnarray}
\frac{d^3\sigma}{dM_{\gamma\gamma} dy d z}
&=& K \Biggr \{  \sum_q 
\frac{1}{48\pi s M_{\gamma\gamma} }\; f_q(x_1) f_{\bar q}(x_2) \,\Biggr [
2 e^4 Q_q^4 \, \frac{1+z^2 }{1-z^2}
+ 2 \pi e^2 Q_q^2 \, M_{\gamma\gamma}^4 \, \eta \, (1+z^2) \,  
+ \frac{\pi^2}{2}\, M_{\gamma\gamma}^8 \, \eta^2\,  (1-z^4) \,  
  \Biggr ] \nonumber \\
&+&
\frac{\pi}{256s}\, f_g(x_1) f_g(x_2) \, M_{\gamma\gamma}^7\eta^2\, 
(1+ 6 z^2 + z^4 ) \; \Biggr \}\;,
\label{diphoton}
\end{eqnarray}
where $z=\cos\theta^*$ is the cosine of the scattering angle in the parton
center-of-mass frame and $y$ is the rapidity of the photon pair.  
For compatibility with the Drell-Yan channel we use the range of $z$ in 
Eq. (\ref{diphoton}) from 
$-1$ to 1, even though the final state photons are indistinguishable from 
each other. We account for NLO QCD corrections via a $K$-factor (see
Eqs. (\ref{dilepton}) and (\ref{diphoton})). We use $K = 1.3$ in the 
calculations. 

For the diagrams with virtual Kaluza-Klein graviton
exchange it is necessary to introduce an explicit upper cut-off, 
of the order of $M_S$, 
to keep the sum over the KK states finite. A naive argument for the
existence of such a cut-off is that the effective theory breaks down
above $M_S$, where detailed understanding of string dynamics is
required.  A recent observation by Bando {\it et al.\/} \cite{bando} suggests
a way around this issue by postulating that the brane is actually ``flexible,''
with a certain tension. When the SM particles that live on the brane
couple to the Kaluza-Klein states of a bulk gravitational field, the
brane has to ``stretch'' out in order to ``catch'' these Kaluza-Klein
states. These stretches are actually quantum fluctuations, usually
suppressed exponentially.  The higher the $n$ of the Kaluza-Klein
state, the stronger the suppression is. From the above argument, the
contribution of high Kaluza-Klein states is suppressed, and the
arbitrary cut-off in the sum over the KK states becomes irrelevant.  
Bando {\it et al.\/} \cite{bando} showed that if the brane tension is 
equal to $M_S$,
total amplitude with such a suppression is the same as
the amplitude on a non-flexible brane with a cut-off scale set at
$M_S$.

Another argument, coming from the fundamental string theory, is that the
coupling constant of each Kaluza-Klein state ($M_k=2\pi k /R$) is, 
in general, not independent of $k$, but exponentially suppressed \cite{anton}:
\begin{equation}
	g(k) \sim  g a(k) \, \exp{ \frac{- c k^2}{R^2 M_S^2} } \;.
\end{equation}
Here $a(k)$ depends on the normalization of the gauge kinetic term and 
$c$ is a constant that depends on the fundamental theory.  Hence, even 
though all the Kaluza-Klein levels are summed over, the sum is not 
divergent and is equivalent to the unsuppressed sum with a certain 
cut-off scale.

Following these recent observations, we relaxed the assumption about
the cut-off scale $\Lambda$ when summing the effects of all virtual
graviton propagators.  In Han {\it et al.}, at amplitude level, the
sum of the propagators $\sum_k i/(\hat s - m_k^2 +i\epsilon)$ is truncated
for $m_k > \Lambda$, where $\Lambda$ is chosen to be $M_S$.  In this work, we
allow $\Lambda$ to be different from $M_S$, but in order for the
effective theory to be valid, it is required that $\Lambda\sim
O(M_S)$.  We define a scale factor $c=O(1)$, such that $\Lambda = c
M_S$.  After this modification, the corresponding change in the above
equations is:
\begin{equation}
\label{ccc}
\eta = \frac{\cal F}{M_S^4} \longrightarrow \frac{{\cal F} c^{n-2}}{M_S^4} \;.
\end{equation}
In the numerical analysis, we use $\eta$ as the fit parameter in order to 
reduce
non-linearity of the problem. Once the best fit value of $\eta$ is obtained,
it is straightforward to obtain the corresponding value of $M_S$ for 
given $n$ and $c$.

\section{Procedures}
\label{sec:method}

\subsection{Experimental Acceptance}

We use typical kinematic and geometrical acceptance of the D\O\
detector to estimate the sensitivity in Run I and Run II of the
Tevatron for both dilepton and diphoton production:
\begin{eqnarray}
|y_i| <1.1 \;\; &{\rm or}&\;\; 1.5 < |y_i| < 2.5\;, 
\nonumber \\
M_{ii} > 50\;{\rm GeV}\,\;\; &{\rm and}& 
p_{T_i} > 25 \;{\rm GeV} \;, \nonumber
\end{eqnarray}
where $i = e,\mu,$ or $\gamma$.  The integrated luminosities used in
our study are 130 pb$^{-1}$, 2 fb$^{-1}$, and 20~fb$^{-1}$ for Run I,
IIa, and IIb, respectively.  In addition to the acceptance losses we
take into account the detector resolution effects, as well as the
longitudinal smearing of the primary interaction vertex.

For the LHC ($\sqrt{s}=14$ TeV $pp$ collision) we use the following 
``typical'' acceptance cuts:
\begin{equation}
|y(i)| < 2.5 \;, \;\; 
M_{ii} > 50\;{\rm GeV}\;, \;\; 
p_{T_i} > 25 \;{\rm GeV} \;, \;\;\;\; (i=e,\mu,\gamma) \;.
\end{equation}

In addition to the acceptance losses, we assume the efficiency of
either dilepton or diphoton reconstruction and identification to be
90\% for the LHC or Run II of the Tevatron, and 80\% for Run I.
\footnote{Dimuon acceptance and efficiency were significantly lower
in Run I, but we deliberately have not done a more realistic
simulation in this case. First, the contribution of the dimuon channel
to the overall sensitivity is very small. Second, a designated data
analysis, which is currently being finalized by the D\O\
Collaboration~\protect\cite{GL-SUSY99} will soon override our
estimates by utilizing real collider data and measured
efficiencies.}
In the case of charged leptons, detection inefficiency
comes from the requirements on consistency of a track in the central
detector with a calorimeter energy deposition (electrons) or a track
in the outer muon detector (muons). For photons the inefficiency
primarily comes from the losses due to photon conversions in the material
in front of the central tracker.

\subsection{Monte Carlo data generation}
\label{sec:MC}

In order to estimate the sensitivity of collider experiments to the
low scale gravity model, we need to generate some ``realistic'' data
sets.  To set limits on the scale $M_S$, we assume that the SM is correct
up to the energies of the Tevatron or the LHC.  We use the SM cross
section of dilepton production (the first line in
Eq. (\ref{dilepton})) to generate a smooth double differential
distribution in $M_{\ell\ell}$ and $z=\cos\theta^*$.  We divide the
$M_{\ell\ell}\times z$ plane into a grid of $20 \times 20\; (50\times 20)$
bins, with $M_{\ell\ell}$ from 0 GeV to 2000 (10000) GeV and $z$ from
$-1$ to 1 for Tevatron (LHC).  For each bin $(i,j)$ of this grid, the
expected number of events, $S^{\rm SM}_{ij}$, is obtained by
multiplying the cross section in this bin by the known integrated
luminosity and efficiency. We further proceed with a Monte Carlo (MC)
{\it gedankenexperiment\/}. For each bin $(i,j)$ we generate a
random number of events, $n_{ij}$, using Poisson statistics with the mean 
$S^{\rm SM}_{ij}$. Similar {\it gedankenexperiment\/} is done for the diphoton
production.  We use the dilepton or diphoton MC data sets generated in
this way to perform the best fit to the low scale gravity model 
(see section~\ref{sec:fit}).
Either of the two channels, or their combination can be used in the fit.
\footnote{Note that combination of the dilepton and
diphoton channels implies combination of the corresponding likelihoods, not the
spectra!}

\subsection{Fitting procedure}
\label{sec:fit}

We extract the lower limit on the gravity scale $M_S$ by fitting the
``data'' obtained in a MC experiment with a sum of the SM background and 
Kaluza-Klein
graviton contribution. We employ both the maximum likelihood method
and pure Bayesian approach with a flat prior probability for
$\eta \ge 0$ and 0 for $\eta < 0$. Since we focus on the number of 
large extra dimensions $\ge 3$,
\footnote{For the case of $n=2$, ${\cal F}$ depends on $\hat s$.  In the next
section, we will also give the results for $n=2$ by estimating the average
$\hat s$ for the gravity term in dilepton and diphoton production.}
the fitting procedure is straightforward,
as the factor ${\cal F}$ can be taken out of the integration over the phase
space.

We generate three templates that describe the cross section
in the case of large extra dimensions. The first one describes the SM
cross section on the rectangular grid described in
section~\ref{sec:MC}. The other two describe terms
proportional to $\eta$ (interference term) and to $\eta^2$
(Kaluza-Klein term), respectively. We then parameterize production cross
section in each bin of the $M \times z$ grid as a bilinear
form in $\eta$:
\begin{equation}
	\sigma = \sigma_{\rm SM} + \sigma_4 \eta + \sigma_8 \eta^2,
\label{eq:template}
\end{equation}
where $\sigma_{\rm SM}$, $\sigma_4$, and $\sigma_8$ are the three
templates described above.

\begin{figure}[th]
\centering
\includegraphics{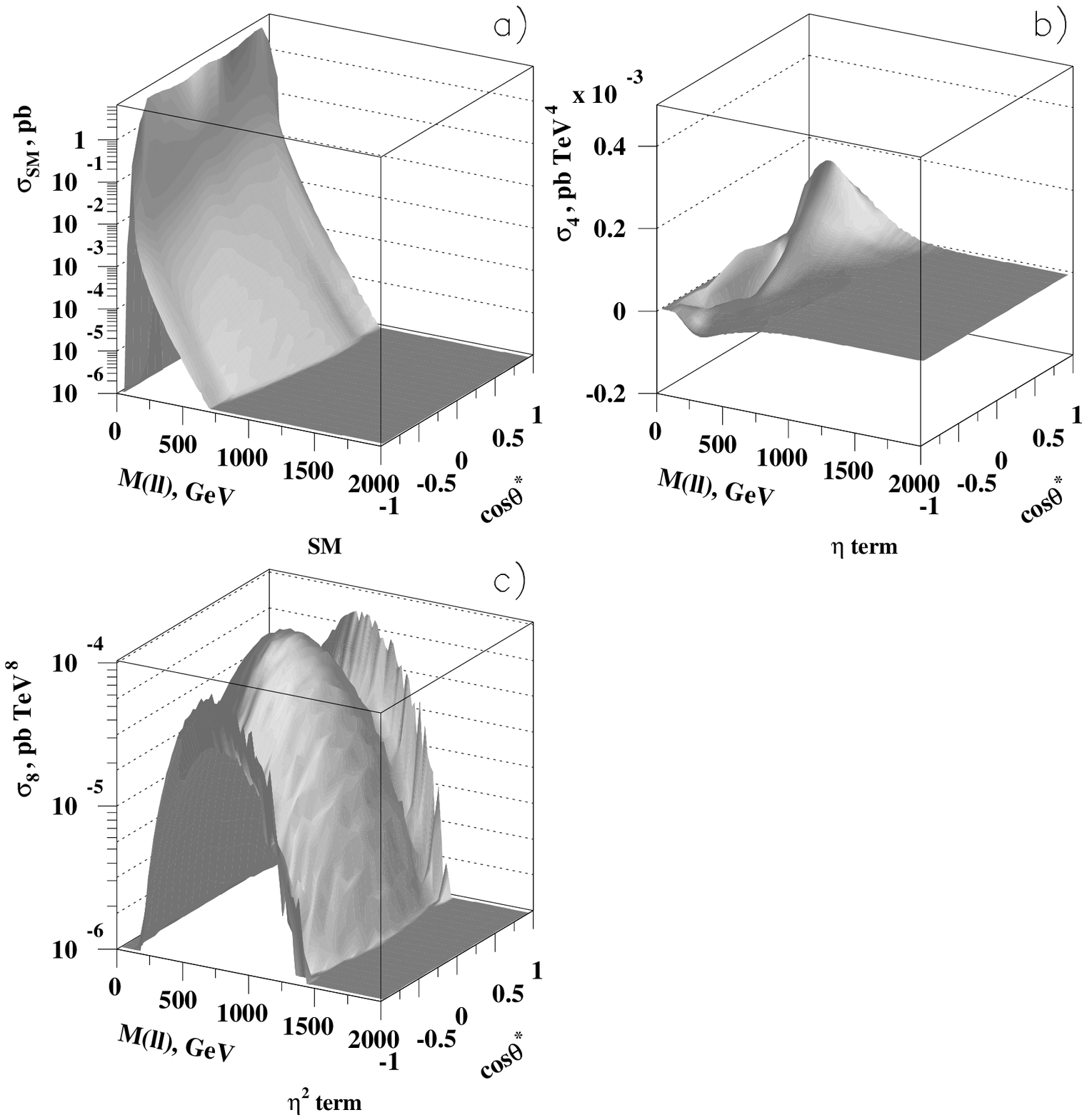}
\caption{\label{fig1}
\small The 3-D plots for double differential distribution $d^2\sigma/
d M_{\ell\ell} d\cos\theta^*$ for Drell-Yan production at the 2 TeV 
Tevatron. (a) SM only, (b) the interference term  
between the SM and the gravity contributions, proportional to $\eta$, 
and (c) the pure gravity 
contribution, proportional to $\eta^2$. Here $\eta={\cal F} c^{n-1}/M_S^4$.
Note that in (b) linear scale is used in the $z$-axis in order to show the
negative $z$ values.}
\end{figure}

\begin{figure}[th]
\centering
\includegraphics{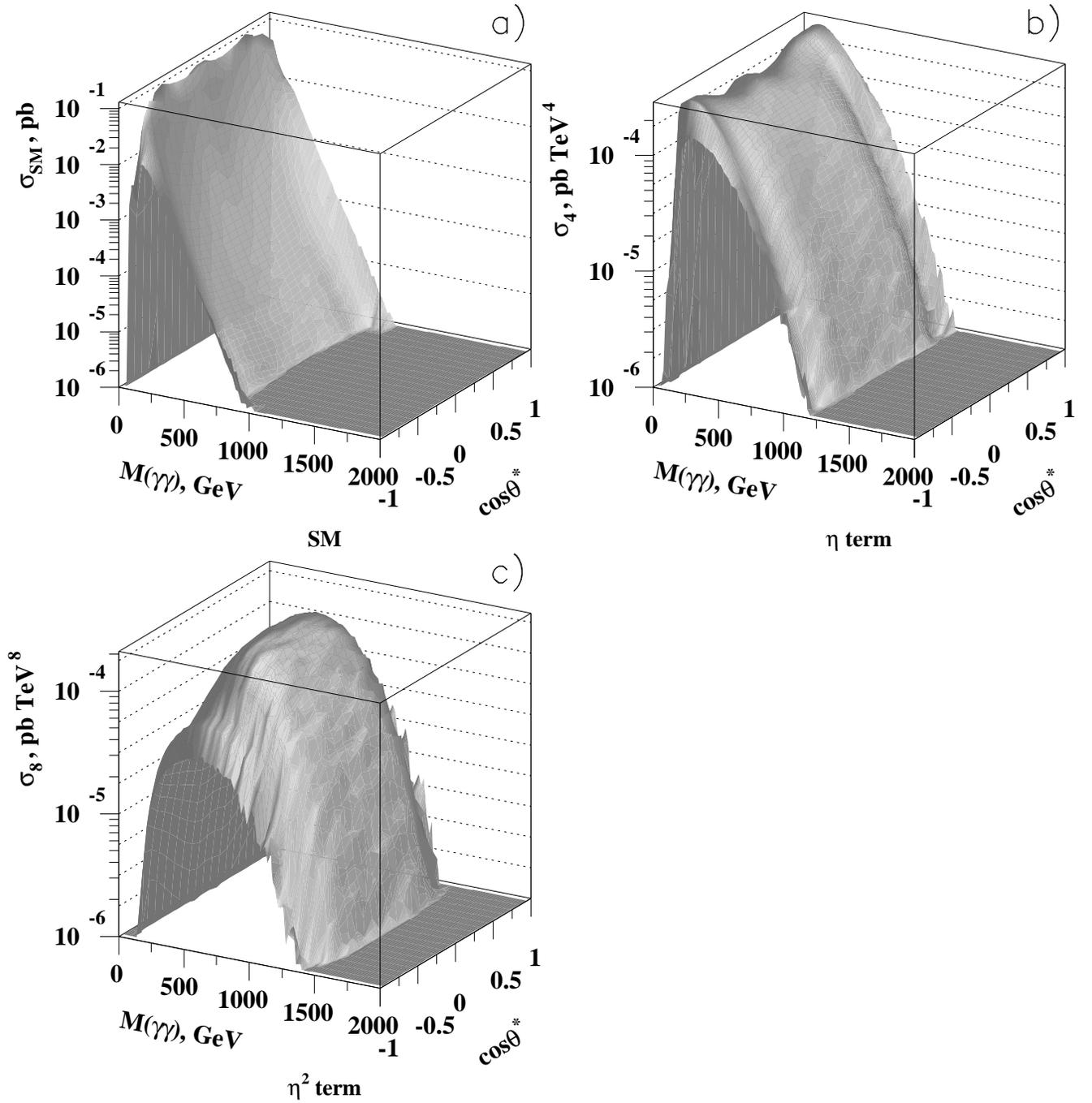}
\caption{\label{fig2}
\small The 3-D plots for double differential distribution $d^2\sigma/
d M_{\gamma\gamma} d\cos\theta^*$ for diphoton production at the 2 TeV
Tevatron.  (a) SM only, (b) the interference terms between the SM and
the gravity contributions, proportional to $\eta$, and (c) the pure
gravity contribution, proportional to $\eta^2$. Here $\eta={\cal F}
c^{n-1}/M_S^4$.}
\end{figure}

In Figs. \ref{fig1} and \ref{fig2}, we show the 3-D plots for the pure
SM, the interference, and pure gravity contributions for dilepton and
diphoton production, respectively, at the 2 TeV Tevatron.  It is clear
that the pure SM decreases rapidly with the invariant mass.  This is
in contrast with the pure gravity contribution that rises quite
sharply with the invariant mass and then turns over due to the effect
of parton distribution functions. The interference term also shows
similar characteristics.  The angular distribution also exhibits
substantial difference among the pure SM, pure gravity, and the
interference. Note the asymmetry of the interference term for dilepton
production (Fig.~\ref{fig1}b) that arises from the charge asymmetry of
the Tevatron beams and final state particles. Analogous distribution
for diphotons or in the LHC case is symmetric.

The probability to observe certain set of data ${\cal N} =
\{n_{ij}\}$, where $(i,j)$ are the bins in $M$ and $\cos\theta^*$,
respectively, as a function of $\eta$ is given by the Poisson
statistics:
\begin{equation}
	P({\cal N}|\eta) = \sum_{ij} \frac{S_{ij}^{n_{ij}}
	e^{-S_{ij}}}{n_{ij}!},
\end{equation}
where $S_{ij} \equiv {\cal L} \epsilon \sigma_{ij}$, and ${\cal L}$ is
the integrated luminosity, $\epsilon$ is the identification
efficiency, and $\sigma_{ij}$ is the cross section given by 
Eq. (\ref{eq:template}), integrated over the bin $(i,j)$.

We now can use Bayes theorem to obtain the probability of $\eta$, 
given the observed set ${\cal N}$:
\begin{equation}
	P(\eta|{\cal N}) = \frac{1}{A} \int dx \;
	\exp\left(\frac{-(x-x_0)^2}{2\sigma_x^2} \right)\;P({\cal N}|\eta),
\end{equation}
where $A$ is the normalization constant, obtained from the unitarity
requirement:
\begin{equation}
	\int_0^\infty \; d\eta \; P(\eta|{\cal N}) = 1\,,
\end{equation}
$x_0$ is the central value of the $\epsilon {\cal L}$, and $\sigma_x$
is the assumed Gaussian error on the quantity $x$. In order to
minimize the uncertainty $\sigma_x$ we perform {\it in situ}
calibration by normalizing $x$ to reproduce the observed number of
events with $M < 100$~GeV (200 GeV) at the Tevatron (LHC) ({\it i.e.\/}, we
use the first mass bin of the MC grid to perform the
normalization). Such a procedure is justified by the fact that
possible contribution from Kaluza-Klein gravitons virtually does not
affect the low mass region (see Figs.~\ref{fig1} and \ref{fig2}). We, 
therefore, assume $\sigma_x$ to be 10\%
or $(100\%/\sqrt{\sum_j n_{1j}})$, whichever is smaller. (When setting
limits on $\eta$ we then only use the mass bins above the normalization
region, {\it i.e.\/} $i > 1$.)

The 95\% C.L. limit on signal, $\eta_{95}$, is
obtained from the following integral equation:
\begin{equation}
\label{eq:e95}
	\int_0^{\eta_{95}} d\eta P(\eta|{\cal N}) = 0.95.
\end{equation}

A less sophisticated likelihood approach does ignore systematic error
on the efficiency and integrated luminosity and simply treats $P({\cal
N}|\eta)$ as the likelihood function. The 95\% C.L. limit in this case is 
obtained by requiring the integral of the likelihood function from the
physics boundary ($\eta = 0$) to $\eta_{95}$ to be equal to 0.95. As
was mentioned before, both approaches yield very close limits on $\eta$.
While the Bayesian technique is a natural way to account for the
systematic errors on the efficiency (and background) estimates (and
this is the approach actually used by the D\O\ experiment to derive
limits), we implemented the classical likelihood approach as well, primarily
to demonstrate the robustness of the limit setting technique.

We further combine the results obtained from the dilepton and diphoton
channels by adding the probabilities (likelihoods) and solving the
integral equation (\ref{eq:e95}) (or its equivalent for the maximum
likelihood method).

As an additional cross check we have tested the fitting techniques
with a set of the MC experiments assuming a non-zero Kaluza-Klein
graviton contribution. Both the Bayesian and maximum likelihood fits
were capable of extracting the input value of the gravity scale
without a systematic bias, as expected.

To convert $\eta_{95}$ from a single MC experiment into a measure of
sensitivity of future experiments, we repeat the above procedures
(both the {\it gedankenexperiment\/} and fit) many times. The limits
obtained in these repeating experiments are histogrammed.  Sensitivity
to the parameter $\eta$ is defined as the median of this histogram,
{\it i.e.\/} the point on the sensitivity curve which 50\% of future
experiments will exceed. All the limits given in the next section are
based on this sensitivity measure. (An alternative approach that
defines sensitivity as the most probable outcome of the {\it
gedankenexperiment\/} agrees with the one we used within 5\%
accuracy.)

\section{Results}
\label{sec:results}

\begin{figure}[th]
\centering
\includegraphics[height=2.8in]{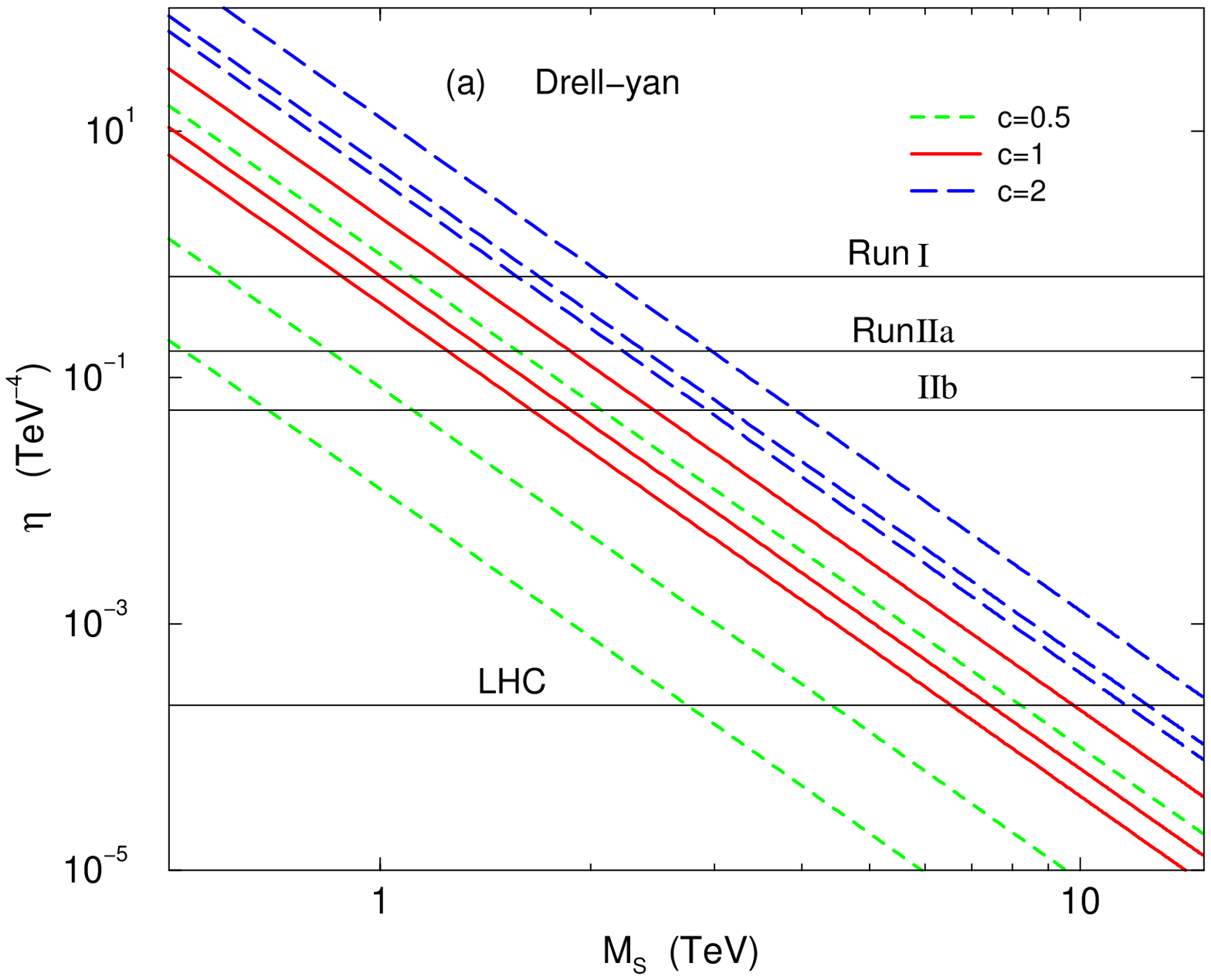}
\includegraphics[height=2.8in]{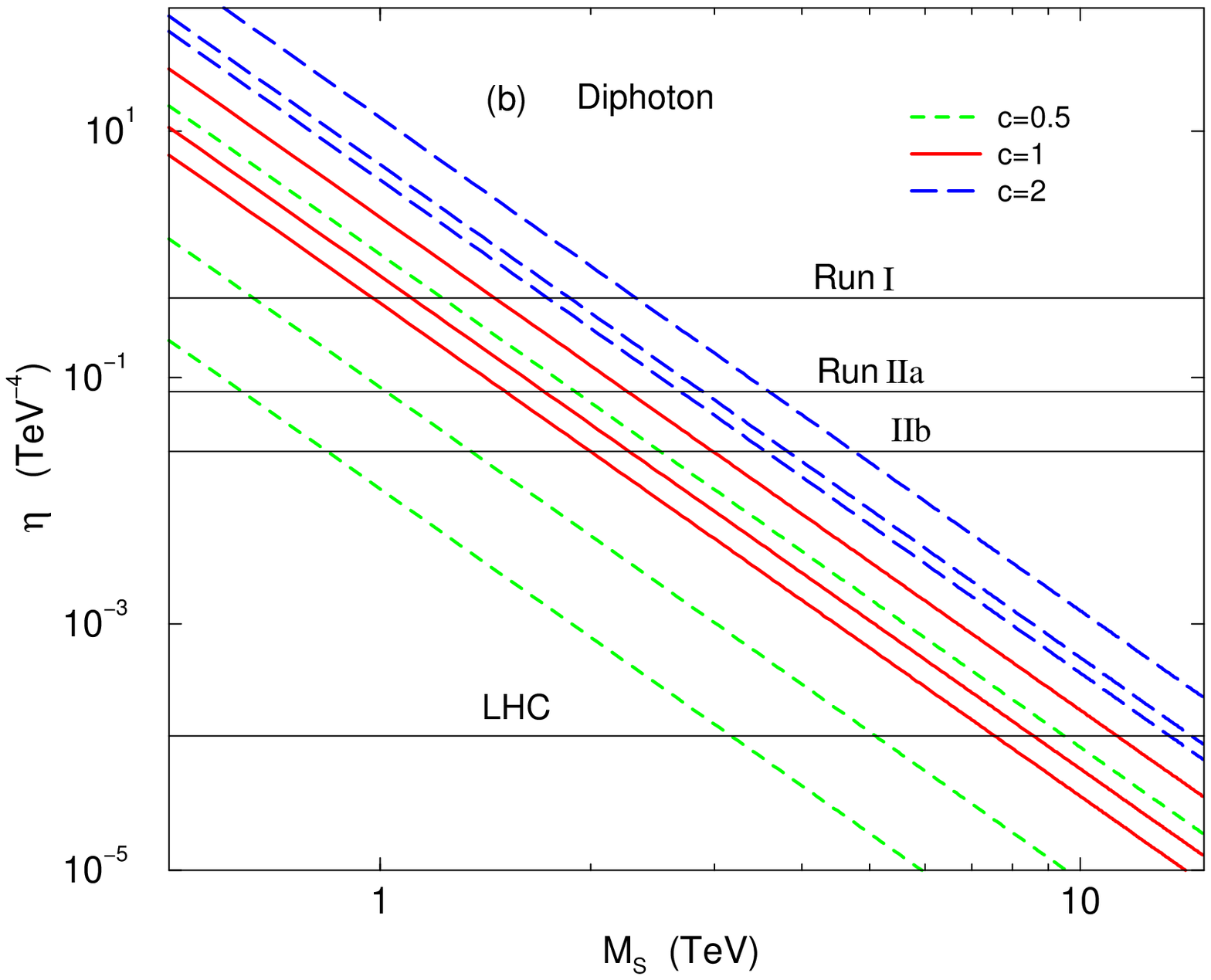}
\includegraphics[height=2.8in]{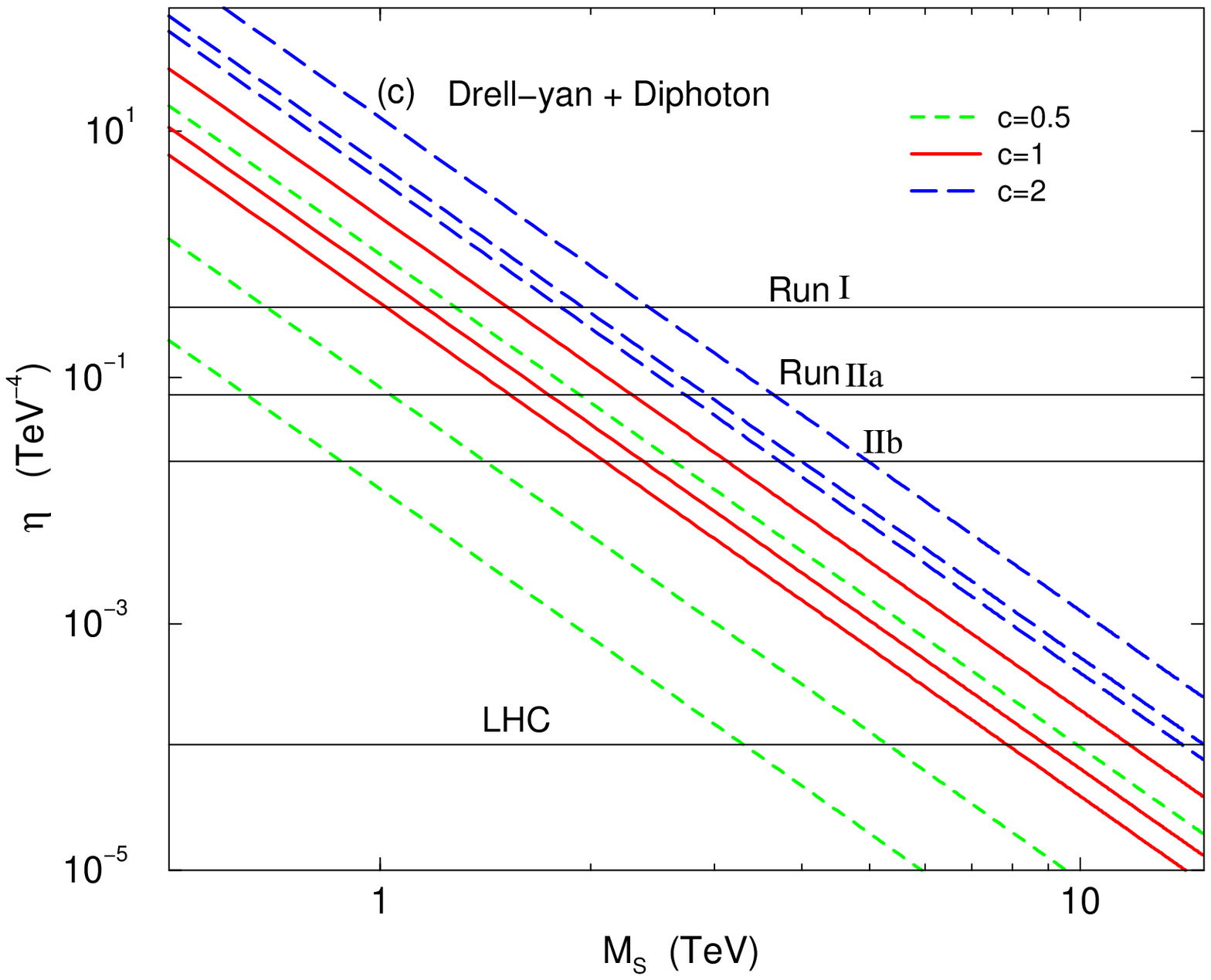}
\caption{\label{fig3}
\small 
$\eta$ versus $M_S$ for a given $(n,c)$.  For $c=0.5,1$ the lines from 
top to bottom are for $n=3,5,7$, whereas for $c=2$ the lines from
top to bottom are for $n=7,5,3$. The 95\% C.L. limit on $\eta$ for
Tevatron Run I, Run II, and for LHC are shown. Limits are based on: 
(a) Drell-Yan production, (b) diphoton production, and (c) combined.}
\end{figure}

In our study, we include both the electron and muon channels in the
Drell-Yan production. In Table~\ref{table1} we show the
sensitivity to $\eta$ in Run I, Run II of the Tevatron, and
at the LHC using dilepton and diphoton production, as well as their 
combination. Corresponding $M_S$ reach is also shown for $n=2-7$ and
$c=1$.  For other values of $c$ they are shown in Fig. \ref{fig3} or
can be calculated by simple rescaling, using Eq. (\ref{ccc}). For the
case $n=2$ the conversion of $\eta$ limits into $M_S$ limits is not
straight-forward, as it depends on the $\hat s$ of the subprocess, see
Eq. (\ref{FF}).  We use the pure gravity contribution in the dilepton and
diphoton production to estimate the corresponding average $\hat
s$. With the average $\hat s$ we can then roughly estimate the $M_S$
limits for $n=2$. For diphoton production the average $\hat s$ for Run
I, Run II, and LHC are (0.61 TeV)$^2$, (0.66 TeV)$^2$, and (3.2
TeV)$^2$, respectively, while for dilepton production the average
$\hat s$ are (0.60 TeV)$^2$, (0.64 TeV)$^2$, and (3.1 TeV)$^2$,
respectively.

\begin{table}[t]
\caption{\small 
Sensitivity to the low scale gravity model parameter 
$\eta={\cal F}c^{n-2}/M_S^4$ 
in Run I, Run II of the Tevatron and at the LHC, using the dilepton, 
diphoton production, and their combination.  The corresponding 
95\% C.L. limits on $M_S$ are given in TeV for $n=2-7$ and $c=1$.  
Results for other $c$ values can be read from Fig. \ref{fig3}
or obtained by rescaling, using Eq. (\ref{ccc}).
\label{table1} }
\medskip
\centering
\begin{tabular}{|c||c|cccccc|}
\hline
& $\eta_{95}$ (TeV$^{-4}$) & $n=2$ & $n=3$ & $n=4$ & $n=5$ & $n=6$ & $n=7$ \\
\hline
\hline
\underline{Run I (130 pb$^{-1}$)}& & & & & & & \\ 
{}Dilepton & 0.66 & 1.21 & 1.32 & 1.11 & 1.00 & 0.93 & 0.88 \\
{}Diphoton & 0.44 & 1.39 & 1.46 & 1.23 & 1.11 & 1.03 & 0.98 \\ 
{}Combined & 0.37 & 1.48 & 1.53 & 1.29 & 1.16 & 1.08 & 1.02 \\
\hline
\underline{Run IIa (2 fb$^{-1}$)} & & & & & & & \\
{}Dilepton & 0.163 & 1.92 & 1.87 & 1.57 & 1.42 & 1.32 & 1.25 \\
{}Diphoton & 0.077 & 2.40 & 2.26 & 1.90 & 1.71 & 1.60 & 1.51 \\
{}Combined & 0.072 & 2.46 & 2.30 & 1.93 & 1.74 & 1.62 & 1.54 \\
\hline
\underline{Run IIb (20 fb$^{-1}$)} & & & & & & & \\
{}Dilepton & 0.054 & 2.70 & 2.47 & 2.08 & 1.88 & 1.75 & 1.65 \\
{}Diphoton & 0.025 & 3.40 & 3.00 & 2.53 & 2.28 & 2.12 & 2.01 \\
{}Combined & 0.021 & 3.54 & 3.11 & 2.61 & 2.36 & 2.20 & 2.08 \\
\hline
\underline{LHC (14 TeV, 100 fb$^{-1}$)} & & & & & & & \\
{}Dilepton & $2.20\times10^{-4}$ & 10.2 & 9.76 & 8.21 & 7.42 & 6.90 & 6.53 \\
{}Diphoton & $1.24\times10^{-4}$ & 12.1 & 11.3 & 9.47 & 8.56 & 7.97 & 7.53 \\
{}Combined & $1.05\times10^{-4}$ & 12.8 & 11.7 & 9.87 & 8.92 & 8.30 & 7.85 \\
\hline
\end{tabular}
\end{table}

The Drell-Yan channel is not as sensitive as the diphoton channel and,
therefore, the combined limit is close to the limit from the diphoton
channel only.  In Run I, using the combination of two channels, the
sensitivity to $M_S$ is about 1.0 to 1.5 TeV for $n=7-2$ and $c=1$.
It increases to 1.5 to 2.5 TeV in Run IIa, and 2.1 to 3.5 TeV in Run
IIb. At the LHC, the sensitivity soars up to $7.9-12.8$ TeV.  Both
higher center-of-mass energy and increase in the integrated luminosity
help to improve the limits.

We also study the improvement in the sensitivity from the double
differential $d^2\sigma /dM d\cos\theta^*$ fit compared to that from
the single differential $d\sigma/dM$ fit.  We have repeated the entire
procedure with a $20\times 1$ grid in the $(M\times \cos\theta^*)$
plane, which is equivalent to fitting the single differential distribution
$d\sigma/dM$.  Corresponding limits in Run IIa deteriorate to:
\begin{eqnarray}
 \eta_{95} = 0.176\;{\rm TeV}^{-4} && \mbox{for dilepton} \;, \\
 \eta_{95} = 0.089\;{\rm TeV}^{-4} && \mbox{for diphoton} \;, \\
 \eta_{95} = 0.084\;{\rm TeV}^{-4} && \mbox{for combined dilepton and diphoton}
\;.
\end{eqnarray}
By using the double differential cross section we achieve an
improvement of about 10\% (15\%) in the limit on $\eta$ for dileptons
(diphotons). While such an improvement in sensitivity translates only
into a few per cent increase in the limit on $M_S$, it is actually
equivalent to a 30\% decrease in the integrated luminosity, required
to set a certain limit on $M_S$.

\section{Conclusions}
\label{sec:conclusions}

The sensitivity to the effective Planck scale $M_S$ obtained in this
analysis supercedes those from the previous studies, in which only
one-dimensional distributions were used ({\it e.g.\/}, Drell-Yan
production \cite{me1}, diboson production \cite{atwood}, diphoton
production \cite{me2,eboli}, dijet production and top pair production
\cite{mathews}).  The recent work by \'{E}boli {\it et al.} \cite{eboli}
that studied diphoton production in the Tevatron Run IIa and at the
LHC quotes 95\% C.L. upper limits on $M_S$ of 1.73 TeV ($n=4$) in Run
IIa and 7.7 TeV ($n=4$) at the LHC.  Our limits exceeds the latter,
partly because we have taken into account the invariant mass and
angular distributions simultaneously, and partly because we do not
impose the unitarity constraint $\sqrt{\hat s} < 0.9 M_S$ and use
a slightly higher efficiency.

As we have mentioned in the Introduction, the invariant mass $M$ and
the central scattering angle $\cos\theta^*$ already span the entire
phase space of a $2\to 2 $ process.  Thus, our fit method gives an
ultimate way of probing the low scale gravity in the virtual graviton 
exchange processes, because all relevant
information is contained in the $(M\times \cos\theta^*)$ plane.
We have shown that the improvement in the limits of $\eta$ from the double
differential $d^2\sigma/dM d\cos\theta^*$ fit over those from the
single differential $d\sigma/dM$ fit is about 15\%, which corresponds
to a 30\% decrease in the integrated luminosity needed to obtain a
certain sensitivity in $M_S$.

To summarize, we have analyzed the double differential distribution in the
invariant mass and scattering angle for dilepton and diphoton
production at hadron colliders.  We have obtained better sensitivity
than previous studies have achieved. Limits that we obtained using the
Bayesian approach and maximum likelihood method are numerically
identical. The expected limits on $M_S$ are: $0.9-1.5$ TeV (Run I),
$1.3-2.5$ TeV (Run IIa), $1.7-3.5$~TeV (Run IIb), and $6.5-12.8$ TeV
(LHC) for $n=7-2$.

\section*{\bf Acknowledgments}
We would like to thank Konstantin Matchev for helpful discussions. This 
research was supported in part by the U.S.~Department of Energy under 
Grants No. DE-FG02-91ER40688 and DE-FG03-91ER40674, and by the Davis 
Institute for High Energy Physics.


\end{document}